# Detector Development for the European XFEL: Requirements and Status

**Andreas Koch, Markus Kuster, Jolanta Sztuk-Dambietz and Monica Turcato**

European XFEL GmbH, Notkestrasse 85, 22607 Hamburg, Germany

E-mail: andreas.koch@xfel.eu

**Abstract**. The variety of applications and especially the unique European XFEL time structure will require adequate instrumentation to be developed to exploit the full potential of the light source. Two-dimensional integrating X-ray detectors with ultra-fast read out up to 4.5 MHz for 1024 x 1024 pixel images are under development for a variety of imaging applications. The actual status of the European XFEL detector development projects is presented. Furthermore, an outlook will be given with respect to detector research and development, performance optimization, integration, and commissioning.

## 1. Introduction

The European X-ray Free-Electron Laser Facility (XFEL.EU) is an international scientific user facility under construction in Hamburg, Germany that will start user operation in 2016 [1]. It is an X-ray photon source providing laterally coherent X-rays for six experimental stations (start-up configuration) in the range of approximately 250 eV to 25 keV. An electron beam is accelerated on a linear path up to energies of 17.5 GeV by superconducting cavities. The electrons then generate coherent X-rays in a series of undulators up to 200 m in length based on the SASE process (Self-Amplified Spontaneous Emission). The electron acceleration, photon beam generation and beam transport path extends over a total length of 3.4 km providing particularly intense and short X-ray pulses down to a few femtoseconds with a peak brilliance of $10^{33}$ photons/s/mm$^2$/mrad$^2$/0.1%BW. The electron source will deliver trains of typically 2700 pulses at 4.5 MHz. Each train will be followed by a gap of 99.4 ms.

The high radiation intensity and short pulse duration will open a window on new scientific techniques in photon science. At the start of the project no imaging detectors were available for the very demanding requirements, especially with respect to the high brilliance and the 4.5 MHz pulse structure. To make optimal use of the capabilities of the XFEL.EU, a development program was launched in 2006 for three detector projects [2] that are managed by different consortia [3]: the DEPFET Sensor with Signal Compression (DSSC) for the SQS (Small Quantum Systems) and SCS (Spectroscopy and Coherent Scattering) experiments at XFEL.EU, the Adaptive Gain Integrating Pixel Detector (AGIPD) for SPB (Diffraction of Single Particles and Biomolecules), MID (Materials Imaging and Dynamics), HED (High Energy Density Matter), and the Large Pixel Detector (LPD) for FXE (Femtosecond X-ray Experiments). In addition, slower detectors based on more readily available technology are also now being developed to mitigate risks for day-one operation. These are currently being adapted to the experimental stations: a modified pnCCD [4] and an FCCD [5]. This article summarises the requirements for detectors at the European XFEL as well as the status and progress of our detector development program. A brief outlook on possible future developments is also given.

## 2. Progress on the detector development program at the European XFEL

The different scientific applications at XFEL.EU will be mainly addressed by the imaging detector development projects shown in table 1. The first column summarises the requirements of the different experimental stations, the other columns the specifications of each detector. In more detail the actual status of the DSSC [6] (hexagonal pixel), the AGIPD and the LPD detector projects are as follows:

**Table 1.** Overview on 2D imaging detector development projects at the European XFEL. All detectors are based on Si sensor technology and will provide a spatial resolution close to the pixel size. All detectors are actively cooled and foreseen for vacuum operation, except the LPD.

|  | Requirements | AGIPD | DSSC | LPD | pnCCD* | FCCD |
|---|---|---|---|---|---|---|
| **Technology** |  | Hybrid pixel | Hybrid pixel | Hybrid pixel | CCD | CCD |
| **Pixel size** | 10...100's µm | 200x200 µm$^2$ | 204x236 µm$^2$ | 500x500 µm$^2$ | 75x75 µm$^2$ | 30x30 µm$^2$ |
| **Detector size** | 1kx1k | 1kx1k | 1kx1k | 1kx1k | 256x256 | 1kx2k |
| **Tiling, hole** | Central, variable hole | Multiple tiles, variable hole | Multiple tiles, variable hole | Multiple tiles, variable hole | Monolithic no hole | Monolithic fixed hole |
| **Quantum efficiency** | >80% | >80% | >80% | >80% | >80% | >80% |
|  |  |  | 0.5-13 keV | 1-13 keV | 0.3-13 keV | 0.3-6 keV |
| **Sensor thickn.** |  | 500 µm | 450 µm | 500 µm | 450 µm | 200 µm |
| **Energy range** | 0.25-25 keV | 3-13 keV | 0.5-25 keV | 1-25 keV | 0.05-20 keV | 0.25-6 keV |
| **Dynamic range** | $10^3 \ldots 10^4 \ldots$ | $10^4$ at 12 keV | $10^4$ | $10^5$ at 12 keV | $10^3$ at 12 keV | $10^3 \ldots 10^4$ |
| **Noise** | Single photon | 300 el. rms | 50 el. rms | 1000 el. rms | 2 el. rms | 25 el. rms |
| **Frame rate** | 4.5 MHz, 2700 images, 10 bursts/s | 4.5 MHz, 352 images analogue on-chip | 4.5 MHz, 640 images digital on-chip | 4.5 MHz, 512 images analogue on-chip | 200 Hz, continuous (*prototype) | 200 Hz (1kx1k), continuous |

DSSC: First experimental results have been obtained by coupling the non-linear DEPFET prototype to an ASIC prototype comprising the complete readout chain from the analog front-end to the ADC and the memory. Measurements show that the non-linear gain curve is well controlled by the applied voltages, which is an important milestone for further calibration studies. Radiation hardness generally is not critical since the device is targeted at low energies where the sensor is shielding the ASIC. Software to model detector response have been developed and tested with experimental data.

AGIPD: Different ASIC versions have allowed improvements to noise and radiation hardness. The droop (leakage) rate of the ASIC memory is a concern. Up to 10 MGy (entrance dose) droop is less than 40% after 100 ms and can be restored to the non-irradiated performance by cooling. In addition, faster readout will be employed to limit storage times to ≈10 ms (i.e. <7% droop at room temperature). Fully functional 16x16 pixel chips bump-bonded to sensors have been produced and characterised. Employing these, a small scientific measurement campaign at DESY's PETRA III synchrotron source is planned for 2012. Guard rings of the sensor have shown to sufficiently improve radiation hardness at the expense of dead border area. Much effort was invested in the development of a simulation tool, HORUS [7] that is also being adapted for other detectors. Measurements on sensor test circuits show that a radiation damage tolerance between 10 - 100 MGy is expected, which is an excellent result.

LPD: A first small 32x128 pixels prototype (1-tile) of the LPD detector has been tested end of 2011 at the DORIS synchrotron at DESY. A new ASIC version is presently under preparation to reduce noise and improve radiation hardness. Radiation hardness is less critical for the LPD detector since an interposer between the sensor and readout ASIC shields critical components such as the memory cells. First scientific measurements using a 2-tile system (64x128 pixels) are planned for 2012.

## 3. Critical detector parameters

Amongst other essential detector parameters, single photon sensitivity and life time dose are discussed here in more detail. Single photon sensitivity is required for most experiments at XFEL.EU, however, not for each image and not for the entire image area. An over-specification may considerably increase costs and delays of a development project. Concerning lifetime dose, liquid scattering experiments have been analysed in detail in order to provide more precise estimations than before [3]. Here, under-estimation of the lifetime dose could result in radiation damage effects in the recorded images and increase costs for repairing sensors, ASIC components and downtime of the experiment.

Other parameters of high risk for the development projects have also been addressed. These include plasma effects that could potentially degrade performance caused by very high local charge deposition from the incoming radiation (e.g. charge collection time, spatial resolution). No significant degradation is expected for the pixel size and frame rates envisaged. However, for pixel sizes smaller than 100 µm the spatial resolution may be effected [8]. Radiation hardness has also been studied extensively on different test objects [9].

Detector calibration, correction of artifacts and stability are currently being addressed by a dedicated working group including all competences from detector development, to scientific applications. In-pixel signal injection circuits are foreseen to support the calibration task.

### 3.1. Single photon sensitivity

Applications that require processing of 2D images with single photon sensitivity are considered here. Their precision of evaluation depends on the signal-to-noise ratio (SNR) and the area or number of pixels in the evaluated image regions. In the presence of Gaussian distributed electronic read noise table 2 relates SNR to detection area. For a single photon signal five times higher than the rms read noise (SNR=5) the probability of a false detection is $P(1|0)=3 \times 10^{-7}$ or one false event in an area of $3 \times 10^6$ pixels.

Simulations of liquid-scattering experiments have shown that an equivalent read noise of 0.2 ph/pixel rms [10]

**Table 2.** Probability of a single false event at single photon detection.

| Probability of a single false detection | Signal-to-noise ratio | Number of pixels |
|---|---|---|
| P(1\|0) | SNR | for P(1\|0) = 1 |
| 1,59E-01 | 1 | 6,29E+00 |
| 2,28E-02 | 2 | 4,38E+01 |
| 1,36E-03 | 3 | 7,35E+02 |
| 3,21E-05 | 4 | 3,11E+04 |
| 2,92E-07 | 5 | 3,42E+06 |

(SNR=5 in table 2) is sufficiently low for most experiments with dilute samples. These results assume averaging of many images over certain annular image regions. Averaging is often required already to reduce photon quantum noise and has the added benefit of reducing the impact of read noise.

### 3.2. Lifetime dose

Lifetime dose requirements have been investigated in more detail for the intensity and energy range ($10^2$ - $10^4$ eV) expected at the XFEL.EU. The damage mechanisms in general are: increased charge density in the $SiO_2$ and the build-up of charges and energy levels (traps) at the $Si$ – $SiO_2$ interface [9]. High-energy physics experiments suffer from different damage mechanisms, mostly bulk or displacement damage. Current experiences from high-energy physics, therefore, can only partly be used for circuit design. Table 3 summarizes the expected lifetime dose of the front-end part of a detector. With the initial assumptions in Table 3 at the full pulse rate approximately 1 GGy year$^{-1}$ can be obtained [3]. However reductions in accumulated yearly dose for a detector are expected, up to a factor of 800 in total: (a) Available beamtime is 50% year$^{-1}$, (b) 50% of beamtime is used for data taking, (c) 50% reduction by an even split of pulses between undulators. (d) Recent results from LCLS show that the maximum signal in liquid scattering is closer to $10^4$ ph/pixel, i.e. a reduction to 20%. (e) The maximum signal will not always affect the same pixels, i.e. a reduction to 50%. (f) The available number of pulses will be lower at low energies. An averaged reduction of pulses down to 10% is estimated. These all reduce the expected yearly absorbed dose to approximately 1 MGy.

This calculation does not account for accidental exposures by the direct beam, strong reflections from recrystallized samples, fluids or borders of sample holders, neither locally higher dose since diffraction spots may concentrate signal to areas smaller than a pixel. The surface dose or shielding inside the circuits will also change with the photon energy range. Radiation hardness up to 10 – 100 MGy has been achieved with the AGIPD prototype. Optimisation of radiation hardness and its calibration is on-going in all projects.

**Table 3.** Lifetime detector dose absorbed in Si in the front-end part of a detector.

| Initial assumptions | Photon energy | 10 keV |
|---|---|---|
| | Detector saturation | $10^5$ ph/pixel/pulse |
| | Pixel size | 500 µm x 500 µm |
| | Full pulse rate | 30 kHz |
| | Reduced pulse rate | 30 kHz/800 average |
| Lifetime dose | Full pulse rate | 0.7 GGy/year |
| | Reduced pulse rate | 0.9 MGy/year |

## 4. Outlook to future developments

When the XFEL.EU detector program was defined in 2006 technology at the time did not permit the fulfillment of all the optimum requirements, i.e. full digital storage of an X-ray pulse train of 2700 images with 4.5 MHz, repeated at 100 ms, with pixels of 100 µm to 50 µm or even smaller. However, since then sensor and ASIC technology have advanced. In addition, there are scientific applications developing that require large angular coverage resulting in a number of pixels that will significantly exceed the 1 Mpixel systems currently in development.

Stacked wafers based on three-dimensional integrated circuits could provide a solution for the first requirement, i.e. small pixels and full XFEL.EU pulse rate support (see [11] for an overview). Large detector surfaces may be achieved at lower costs with organic electronics where vacuum deposition can be replaced by printing technologies. Additionally, surfaces from organic electronics may be flexible that can be formed into three dimensional shapes, which is interesting for some diffraction experiments. Even though radiation hardness and electronic properties like charge-carrier mobility are not yet at the level of crystalline silicon, the performance of such technology is rapidly increasing [12].

## 5. Conclusions

Detector development at European XFEL for the full frame rate of 4.5 MHz is supported by three important development consortia. Two additional detector activities have also started recently: a pnCCD and FCCD. The projects are progressing well, and the first beamline tests on assembled prototypes are planned for the end of 2012. A second phase of detector developments in the near future could address the capture of images with the full XFEL.EU pulse rate with smaller pixels and larger angular coverage. Examples for promising new technologies have been given, e.g. vertical integrated circuits and organic electronics.

**Acknowledgements**
We thank all members of the detector consortia DSSC, AGIPD and LPD for their efforts in the XFEL.EU detector program. We are grateful to K. Haldrup and colleagues from the Center for Molecular Movies, Copenhagen for providing simulation results on liquid scattering experiments and to U. Trunk and H. Graafsma, DESY for their support during the radiation hardness tests.